%% file: paper.tex
\RequirePackage{lineno}
\documentclass[aps,prl,twocolumn,superscriptaddress,showpacs,preprintnumbers,amsmath,amssymb]{revtex4}

\usepackage{afterpage,rotating}
\usepackage{currvita}

\usepackage{array}
\usepackage{siunitx}

\usepackage{tabularx}
\usepackage{nonfloat}
\usepackage{graphicx}
\usepackage{dcolumn}
\usepackage{bm}
\usepackage{subfigure}
\usepackage{mathrsfs}
\usepackage{amsmath,amsbsy,amssymb,lmodern}
\usepackage{epsfig,color}
\usepackage{placeins}
\usepackage{rccol}
\usepackage{comment}
\usepackage{hyperref} 
\usepackage{setspace} 
\usepackage{tabularx} 
\usepackage{multirow}
\usepackage{braket}
\usepackage{accents}

\DeclareRobustCommand\mybar[1]{\accentset{\rule{0.6em}{0.6pt}}{#1}}

\graphicspath{{ps}}

\newcommand{\bepkk}{B_s^0 \rightarrow \eta^\prime K_S^0}
\newcommand{\bdepkk}{B^0 \rightarrow \eta^\prime K_S^0}

\newcommand{\fbssbssall}{f_{B_{s}^{(*)0} \mybar{B}_s^{(*)0}}}

\renewcommand{\arraystretch}{1.1}

\begin{document}



\title{ \quad\\[1.0cm] Search for the decay $\bepkk$}

\input{pub605.tex}

\begin{abstract}
We report the results of the first search for the decay $\bepkk$ 
using $121.4\,{\rm fb}^{-1}$ of data collected at the $\Upsilon(5S)$ resonance 
with the Belle detector at the KEKB asymmetric-energy $e^+ e^-$ collider. 
We observe no signal and set a 90\% confidence-level upper limit of $8.16 \times 10^{-6}$ 
on the $B_s^0 \rightarrow \eta^\prime K_S^0$ branching fraction.
\end{abstract}


\maketitle

\tighten

{\renewcommand{\thefootnote}{\fnsymbol{footnote}}}
\setcounter{footnote}{0}


The measurements of rare decays of hadrons containing the heavy $b$ quark 
provide an indirect way to search for new hypothetical particles 
(see, e.g., Section~17.4 in \cite{Bevan:2014iga}) 
and, 
generally, 
effects not described by the Standard Model (SM). 
In this Letter we describe the first search for the decay $\bepkk$, 
a charmless decay with contributions from 
gluonic and electroweak penguin amplitudes. 
On the one hand, 
processes that include such amplitudes are sensitive to beyond-the-SM physics, 
which could affect decay rates and CP asymmetries~\cite{theoryandexp}. 
On the other hand, even the SM-based theoretical predictions~\cite{bf1,bf2,bf3,bf4,bf5} for the decay $\bepkk$ 
vary between $0.72 \times 10^{-6}$ and $4.5 \times 10^{-6}$, which makes measuring the branching fraction 
for the studied decay valuable in its own right. 

The two-body decay searched for in the analysis described in this Letter 
is also interesting because it includes $\eta^\prime$, 
the particle whose anomalous production 
in inclusive and exclusive $B$ decays, 
first observed by the CLEO experiment more than two decades ago~\cite{CLEO:1998noh,CLEO:2003iqk}, 
became the catalyst for a large body of dedicated theoretical work~\cite{Du:1999iu}, 
followed by a recent experimental study 
of $B_s^0 \to \eta^\prime X_{s\bar{s}}$ 
at Belle 
using a semi-inclusive method~\cite{Belle:2021ylw}. 
While the large rate for exclusive decays, 
such as $B^{\pm} \to \eta^\prime K^\pm$, 
could be accounted for 
by SM factorization~\cite{Kagan:1997qn}, 
any process involving $\eta^\prime$, 
such as the decays of $B_s^0$ mesons, 
could provide valuable information 
about the role of this particle 
in decays of heavy flavors and 
has been an important part of motivation 
for the work presented here. 


The search for the decay $\bepkk$ described in this Letter is based on a data sample of 
$121.4 {\rm \, fb^{-1}}$ collected  with the Belle detector at the KEKB asymmetric-energy 
$e^+e^-$ collider~\cite{KEKB} when it operated at the $\Upsilon(5S)$ resonance. 
The Belle detector is a large-solid-angle magnetic 
spectrometer that consists of a silicon vertex detector,
a 50-layer central drift chamber, an array of
aerogel threshold Cherenkov counters,  
a barrel-like arrangement of time-of-flight scintillation counters, 
and a CsI(Tl) crystal-based electromagnetic calorimeter (ECL) 
located inside a super-conducting solenoid coil that provided a 1.5~T magnetic field.  
An iron flux-return located outside of the coil 
is instrumented to detect $K_L^0$ mesons and to identify muons. 
The detailed description of the Belle detector could be found elsewhere~\cite{Belle}. 

There are three two-body decays of $\Upsilon(5S)$ that serve as sources of $B_s^0$ mesons: 
$B_s^{*0} \mybar{B}_s^{*0}$, $B_s^{*0} \mybar{B}_s^0$ or 
$B_s^0 \mybar{B}_s^{*0}$, and $ B_s^0 \mybar{B}_s^0$. 
The first two channels have relative fractions of  
${\it{f}}_{B_s^{*0} \mybar{B}_s^{*0}}$ = ($87.0 \pm 1.7$)\% 
and 
${\it{f}}_{B_s^{0} \mybar{B}_s^{*0}}$ = ($7.3 \pm 1.4$)\%~\cite{fraction}. 
We reconstruct signal $B_s^0$ mesons coming directly from $\Upsilon(5S)$ decay 
or from the radiative decay of the excited vector state $B_s^{*0}$ 
(the charge-conjugate decay mode is included throughout this Letter). 
The $\Upsilon(5S)$ resonance production cross section 
is $340 \pm 16$~pb~\cite{fraction},  
and $f_s$, its total branching fraction 
for decays to $B_s^{(*)0} \mybar{B}_s^{(*)0}$, 
is $0.201\pm 0.031$~\cite{PDG}. 
Therefore, the Belle data sample is estimated to contain 
($16.60 \pm 2.68) \times 10^6$~$B_s^0$ mesons. 
We obtain the results for the branching fraction $\mathcal{B}(\bepkk)$ 
as well as for the product $f_s \times \mathcal{B}(\bepkk)$. 

We use 
Monte Carlo (MC) generator {\sc EvtGen}~\cite{EvtGen} 
to simulate the production and decay processes, 
and 
{\sc GEANT} toolkit~\cite{GEANT} 
to model detector response. 
To validate our analysis methods and to calibrate parameters of signal probability density function (PDF), 
we use a control sample of two-body decays $B^0 \rightarrow \eta^{\prime} K_S^0$ 
reconstructed in $711 {\rm \, fb^{-1}}$ of $\Upsilon(4S)$ data collected by Belle at $\Upsilon(4S)$.  


To search for $\bepkk$ decay, we first reconstruct $K_S^0 \to \pi^+\pi^-$ 
and $\eta^{\prime} \to \eta \pi^+\pi^-$ followed by the decay $\eta \to \gamma\gamma$.  
For charged pions from $\eta^\prime$ decay 
we require the distance of closest approach to the interaction point 
to be less than 4~cm along the $z$ axis and less than 0.3~cm in the direction perpendicular to it, 
where the $z$ axis is opposite to the direction of the $e^+$ beam. 
Transverse momenta of these charged pion candidates are required to exceed 100~MeV/{\it c}. 
To distinguish between charged pions and other particles, 
we apply requirements on the likelihood ratio, 
$R_{h/\pi} = L_\pi / (L_\pi + L_h)$, 
which is based on particle identification (PID, see Chapter~5 in \cite{Bevan:2014iga}) measurements, 
where $L_\pi$ is the likelihood for the track according to pion hypothesis, 
and $L_h$ is for kaon ($h=K$) or electron ($h=e$) hypotheses. 
By requiring $R_{K/\pi} \leq 0.6$ and $R_{e/\pi} \leq 0.95$ for charged pion candidates 
we reject 25.8\% of background events while the signal efficiency loss is 7.6\%. 
Photons are reconstructed as ECL energy clusters not matched to any charged tracks. 
We require the reconstructed laboratory-frame energy of photon candidates 
in the barrel (endcap) region of ECL to exceed 50 (100) MeV. 
Barrel region of ECL covers polar angle $\theta$ between $32.2^\circ$ and $128.7^\circ$, 
where the angle $\theta$ is measured w.r.t. the $z$ axis in the laboratory frame. 
$\theta$ coverage of forward and backward endcaps is 
between $12.0^\circ$ and $31.4^\circ$, and $131.5^\circ$ and $157.2^\circ$, respectively. 
The $\eta$ candidates are reconstructed using the decay channel $\eta\to\gamma\gamma$, 
with the reconstructed invariant mass of each candidate required to be 
between $0.515 {\, \rm GeV/{\it c}^2}$ and $0.580 \, {\rm GeV/{\it c}^2}$,
which corresponds, approximately, to a $\pm3\sigma$ 
Gaussian resolution window 
around the nominal $\eta$ mass~\cite{PDG}. 
To suppress 
combinatorial background arising due to low-energy photons, 
the magnitude of the cosine of the helicity angle 
($\cos\theta_{\textrm{hel}}$) 
is required to be less than 0.97, 
where 
$\theta_{\textrm{hel}}$ is the angle 
in the rest frame of the $\eta$ candidate 
between the directions 
of its Lorentz boost from the laboratory frame 
and one of the photons. 
This requirement rejects 11.4\% of background events 
while the efficiency loss for signal events is 3.0\%. 
We perform kinematic fits 
constraining the reconstructed masses 
of the $\eta$ candidates 
to the nominal $\eta$ mass~\cite{PDG}. 
Then $\eta^\prime$ candidates are reconstructed using $\eta$ candidates 
and pairs of oppositely charged tracks identified as pions 
within a wide window of the reconstructed invariant mass $M(\pi^+\pi^-\eta)$ 
between $0.920 {\, \rm GeV/{\it c}^2}$ and $0.980 {\, \rm GeV/{\it c}^2}$, 
which corresponds,
approximately, 
to the range $[-10,+6]\sigma$ 
of the Gaussian resolution 
and includes a wide sideband, 
so $M(\pi^+\pi^-\eta)$ 
could be used to extract the signal, 
as described later in this Letter. 
To identify the $K_S^0$ candidates, 
we use a neural network technique~\cite{NN} 
to search for secondary vertices 
associated with pairs of oppositely charged tracks treated as pions~\cite{ks_reco}. 
%
%
To improve mass resolution, a kinematic fit is performed to the vertex. 
To reconstruct a $B_s^0$ candidate, 
we combine $K_S^0$ and $\eta^\prime$ candidates 
after constraining the reconstructed mass 
of the $\eta^\prime$ to the nominal $\eta^\prime$ mass~\cite{PDG}. 
We further select $B_s^0$ candidates using three kinematic variables: 
beam-energy-constrained $B_s^0$ mass $M_{\rm bc} = \sqrt{E_{\rm beam}^2-p_{B_s^0}^2}$, 
the energy difference $\Delta E = E_{B_s^0}-E_{\rm beam}$, 
and 
$M(\pi^+\pi^-\eta)$, 
where $E_{\rm beam}$ is the beam energy, 
and $E_{B_s^0}$ and $p_{B_s^0}$
are the reconstructed energy and momentum of the $B_s^0$ candidate, respectively, 
calculated in the $e^+ e^- $ center-of-mass frame. 
Signal $B_s^0$ candidates are required to satisfy 
$5.300 {\, \rm GeV/{\it c}^2} < \, M_{\rm bc}< \, 5.440 {\, \rm GeV/{\it c}^2}$ 
and 
$-0.400 {\, \rm GeV} < \, \Delta E < \, 0.300 {\, \rm GeV}$.

The main source of background to our signal is hadronic continuum, 
i.e., quark-pair production in $e^+ e^-$ annihilation. 
To suppress continuum background, 
we take advantage of the difference between 
signal and background event topologies by 
utilizing a set of 17 modified Fox-Wolfram moments~\cite{SFW}. 
By optimizing Fisher discriminant~\cite{fisher} coefficients evaluated using these moments, 
we calculate a likelihood ratio ($\mathcal{R}_{s/b}$) 
according to signal and background hypotheses. 
To suppress background, we require $\mathcal{R}_{s/b}>0.6$. 
This 80.5\%-efficient requirement removes 90.0\% of continuum background.  
The details of our continuum suppression algorithm are provided elsewhere~\cite{KSFW}. 

After applying the described selection criteria, 
16\% of signal MC events have more than one candidate. 
We select the best candidate with the smallest value of 
$\chi^2 = \chi_\eta^2 + \chi_{\pi^+\pi^-}^2 + \chi_{\eta^\prime :\pi^+ \pi^-}^2$, 
where 
$\chi_\eta^2 = \Big(\frac{M_{\gamma\gamma}-m_{\eta}}{\sigma_{\gamma\gamma}}\Big)^2$ 
is from the kinematic fit for the $\eta$ candidate, 
$\chi_{\pi^+\pi^-}^2$ is from the vertex fit for pion candidates from the $K_S^0$ decay, 
and 
$\chi_{\eta^\prime : \pi^+ \pi^-}$ is from fitting charged pion tracks from $\eta^\prime$ decay to a common vertex. 
This method chooses the correct $B_s^0$ candidate 91\% of the time in signal MC events. 
The overall reconstruction efficiency in this analysis is 26.8\%.

To extract the signal yield, 
we perform a three-dimensional (3D) unbinned extended maximum likelihood (ML) fit 
to 
$M_{\rm bc}$, $\Delta E$, and $M(\pi^+\pi^-\eta)$. 
The likelihood function is defined as

\begin{equation}
\mathcal{L}=\frac{e^{-\sum_{j}^{b,s} N_j}}{N!}\prod_{i=1}^{N}\left(\sum_{j}^{b,s} N_{j}\mathcal{P}_{j}[M_{\rm bc}^i, \Delta E^i, M^i(\pi^+\pi^-\eta) ]\right), 
\end{equation}

\noindent where $N$ is the total number of events in the sample, 
$N_j$ are the fit parameters for the number of signal 
($j=s$) 
and background events ($j=b$), 
$\mathcal{P}_j$ are the PDFs 
for the signal and background components of our fitting model. 
The background PDF is further represented by the sum of two 3D PDFs 
which describe a peaking $M^i(\pi^+\pi^-\eta)$ component  
with real $\eta^\prime$ mesons 
and a non-peaking component of combinatorial origin. 
Since the correlations among these three fitting variables are small, 
each of the three 3D PDFs describing the signal contribution, 
and the peaking and non-peaking backgrounds, 
is assumed to factorize as 

\begin{multline}
\mathcal{P}_j[M_{\rm bc},\Delta E, M(\pi^+\pi^-\eta)] = \\
\mathcal{P}_j[M_{\rm bc}] 
\times 
\mathcal{P}_j[\Delta E]
\times  
\mathcal{P}_j[M(\pi^+\pi^-\eta)]. 
\label{eqn:factor}
\end{multline}

The signal component is further described by the sum of contributions from three signal sources 
$B_s^{*0} \mybar{B}_s^{*0}$, $B_s^{*0} \mybar{B}_s^0$ or $B_s^0 \mybar{B}_s^{*0}$, and $ B_s^0 \mybar{B}_s^0$, 
with relative fractions for the two former contributions according to their branching fractions~\cite{fraction}. 

The signal $M_{\rm bc}$ distribution is modeled with a Gaussian, 
and that of $\Delta E$ by the sum of a Gaussian and Crystal Ball function~\cite{CB} (with different means). 
A sum of two Gaussians with the same mean is used to describe 
the reconstructed invariant mass of $\eta^\prime$ candidates in signal events.

To model the background $M_{\rm bc}$ distribution, an ARGUS~\cite{ARGUS} function 
with a fixed endpoint at $5.433 \, \rm GeV/{\it c}^2$ is used. 
We use a second-order Chebyshev polynomial to describe the background $\Delta E$ distribution. 
To account for the presence of real $\eta^\prime$ mesons in background events, 
we use the signal $M(\pi^+\pi^-\eta)$ PDF to model the peaking part 
and 
a first-order Chebyshev polynomial to model the non-peaking component.

To obtain PDF shape parameters for signal, 
we first use the $\Upsilon(5S)$ signal MC sample 
and determine the peak positions for $M_{\rm bc}$ and $\Delta E$. 
Then we use $\Upsilon(4S)$ data for the decay $B^0 \rightarrow \eta^\prime K^0_S$ 
to determine all the other PDF parameters. 
To obtain background PDF shapes, 
we use $\Upsilon(5S)$ sideband data 
collected outside of the signal region defined as 
$5.401 {\, \rm GeV/{\it c}^2} < \, M_{\rm bc}< \, 5.423 {\, \rm GeV/{\it c}^2}$ 
and 
$-0.200 {\, \rm GeV} < \, \Delta E < \, 0.100 {\, \rm GeV}$, 
and
$0.940 {\, \rm GeV/{\it c}^2} < \, M(\pi^+\pi^-\eta) \, < 0.970 {\, \rm GeV/{\it c}^2}$. 

\begin{figure*}
\small
 \begin{center}
    \subfigure{\includegraphics[width=.99\textwidth]{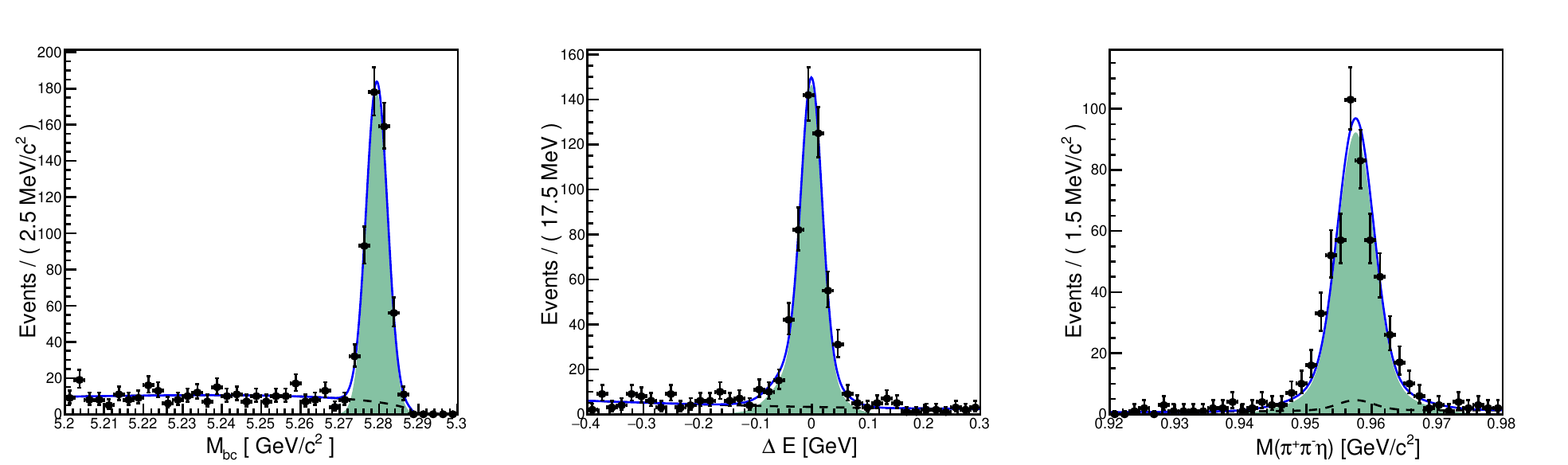}}
 \end{center}
\figcaption{Signal region fit projections onto $M_{\rm bc}$, $\Delta E$ and $M(\pi^+\pi^-\eta)$ 
for $\bdepkk$ event candidates in $\Upsilon(4S)$ data 
after additional selection criteria are applied, as described in the text. 
Points with the error bars show the binned data. 
Blue solid lines show the results of the fit, 
filled area and black dashed line show the signal and background fit components, respectively.}
\label{fig:Y4S_fitting_results}
\end{figure*}

To validate our $\Upsilon(5S)$ analysis, we use the full Belle data sample collected at $\Upsilon(4S)$ energy 
to analyze the decay $B^0 \rightarrow \eta^\prime K^0_S$. 
The results of the fit to $\Upsilon(4S)$ data 
are shown in Fig.~\ref{fig:Y4S_fitting_results}, 
where each fit projection is plotted after additional selection criteria are applied to the other two variables,  
$0.948 {\, \rm GeV/{\it c}^2} < \, M(\pi^+\pi^-\eta) \, < 0.966 {\, \rm GeV/{\it c}^2}$, 
$5.274 {\, \rm GeV/{\it c}^2} < \, M_{\rm bc}< \, 5.286 {\, \rm GeV/{\it c}^2}$, 
and 
$-0.100 {\, \rm GeV} < \, \Delta E < \, 0.060 {\, \rm GeV}$. 
We estimate the branching fraction, 
${\mathcal{B}}(B^0 \rightarrow \eta^\prime K^0) = (52.3 \pm 2.1) \times 10^{-6}$ 
(where only the statistical uncertainty is shown), 
which is consistent with our previous result~\cite{Belle:2006lqc} 
within the estimated systematic uncertainties. 


To extract the signal yield at $\Upsilon(5S)$, 
we fix all PDF shape parameters to the values 
obtained from our MC-assisted data-based studies, 
except for the fraction of background 
containing real $\eta^\prime$ mesons, 
which remains a free parameter in our final fit. 
To obtain our nominal result, we fit the data 
with the following three floating parameters in the fit: 
the number of signal events $N_{s}$, 
the number of background events 
$N_{b}$, 
and the fraction of background with real $\eta^\prime$ mesons. 

\begin{figure*}
\small
 \begin{center}
    \subfigure{\includegraphics[width=.99\textwidth]{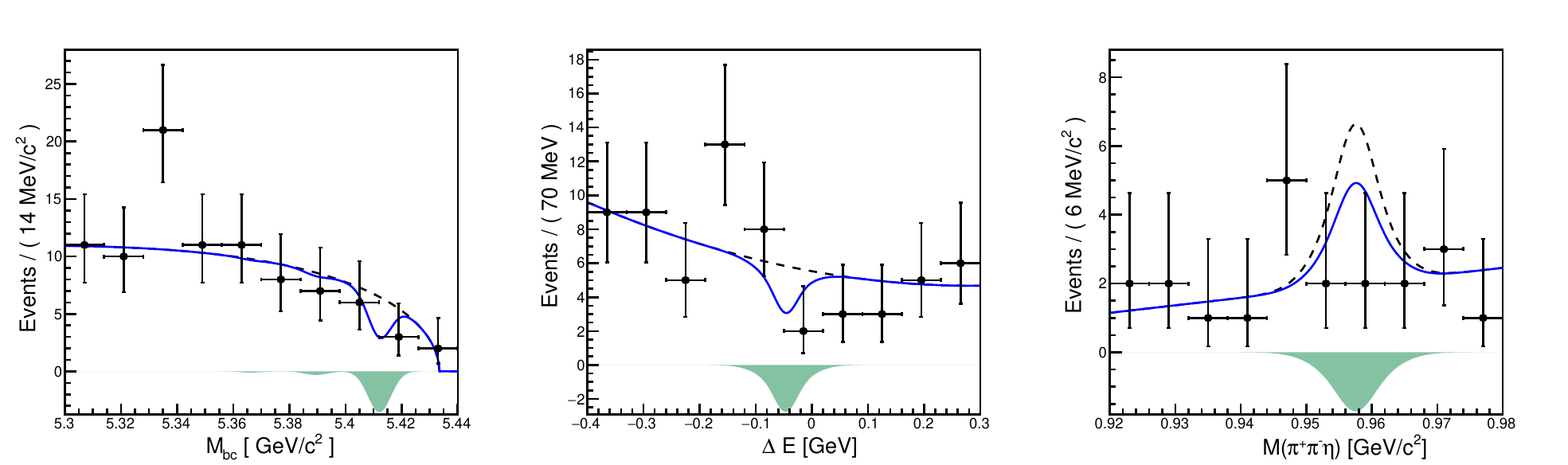}}
 \end{center}
\figcaption{Signal region fit projections onto $M_{\rm bc}$, $\Delta E$ and $M(\pi^+\pi^-\eta)$ 
for $\bepkk$ event candidates in $\Upsilon(5S)$ data 
after additional selection criteria are applied, as described in the text. 
Points with the error bars show the binned data. 
Blue solid lines show the results of the fit, 
filled area and black dashed line show the signal and background fit components, respectively.}
\label{fig:fitting_results}
\end{figure*}

By performing a 3D fit to $\Upsilon(5S)$ data, 
we obtain $-3.21 \pm 1.85$ signal and $801 \pm 28$ background events. 
The results of the fit are plotted in Fig.~\ref{fig:fitting_results}. 
To emphasize the dominant signal source, $B_s^{*0} \mybar{B}_s^{*0}$, 
each fit projection in this figure is plotted after additional selection criteria are applied to the other two variables,  
$0.948 {\, \rm GeV/{\it c}^2} < \, M(\pi^+\pi^-\eta) \, < 0.966 {\, \rm GeV/{\it c}^2}$, 
$5.400 {\, \rm GeV/{\it c}^2} < \, M_{\rm bc}< \, 5.440 {\, \rm GeV/{\it c}^2}$, 
and 
$-0.100 {\, \rm GeV} < \, \Delta E < \, 0.060 {\, \rm GeV}$. 
We observe no signal and estimate the upper limits 
for the branching fraction and its product with $f_s$.

Sources of systematic uncertainties and their contributions are summarized in Table~\ref{tab:systematics}. 
The relative uncertainties for $f_s$ and $\sigma(\Upsilon(5S))$ are 15.4\%~\cite{PDG} and 4.7\%~\cite{fraction}, 
respectively. 
Systematic uncertainty due to $\fbssbssall$, 
i.e., relative contributions of the three signal sources, 
is 1.87\%, 
estimated by varying the relative fractions of the three contributions to signal PDF. 
For daughter branching fractions, the uncertainties for 
$\eta \rightarrow \gamma\gamma$, $\eta^\prime \rightarrow \eta\pi^+\pi^-$, and $K_S^0 \rightarrow \pi^+ \pi^-$ 
are 
0.2\%, 0.7\%, and 0.05\%, respectively~\cite{PDG}. 
Statistical uncertainty due to MC statistics is 0.11\% 
via $\sqrt{\epsilon \times (1-\epsilon)/N}$, 
where $N$ is the total number of signal MC events, 
and $\epsilon$ is the overall reconstruction efficiency. 
The uncertainties in 
$\pi, \, \eta$, and $K_S^0$ 
reconstruction efficiencies 
are 
1.4\% (0.35\% per track~\cite{track_syst}), 
4.1\%~\cite{etasystematics}, 
and 
1.4\%~\cite{ks_syst}
per particle, respectively. 
The uncertainty due to PDF parametrization is 11.9\%, 
estimated from the change in signal yield 
while varying fixed PDF shape parameters 
one at a time by one unit of their Gaussian uncertainties 
as measured from the control data sample for signal 
and from data sideband for background. 

The uncertainty from PID selection is 2.4\%~\cite{track_syst}. 
By comparing the $\mathcal{R}_{s/b}$ distributions for 
$B_s^0 \rightarrow K_S^0 \eta^{\prime}$ 
events 
in $\Upsilon(4S)$ data and signal MC events, 
we estimate the uncertainty in the efficiency 
of likelihood ratio requirement to be 4.4\%. 
We estimate the total multiplicative uncertainties to be 
17.6\% for $ f_s \times \mathcal{B}(B_s^0 \rightarrow K_S^0 \eta^{\prime})$ 
and 
23.4\% for $\mathcal{B}(B_s^0 \rightarrow K_S^0 \eta^{\prime})$. 

\begin{table}[!htb]
\centering
\caption{Summary of relative systematic uncertainties for 
$\mathcal{B}(\bepkk)$ 
and 
$f_s \times \mathcal{B}(\bepkk)$.}
\bgroup
\def\arraystretch{1.25}
\begin{tabular}{lS}
\hline \hline
\multicolumn{1}{c}{\rm Source} & {\rm Uncertainty (\%)}  \\ 
\hline 
$\sigma(\Upsilon(5S))$          & 4.7  \\
$\fbssbssall$                          & 1.87  \\
$\mathcal{B}(\eta \rightarrow \gamma\gamma)$ & 0.2  \\
$\mathcal{B}(\eta^\prime \rightarrow \eta \pi^+ \pi^-)$  & 0.7\\
$\mathcal{B}(K_S^0 \rightarrow \pi^+ \pi^-)$  &0.05 \\
MC statistics                   & 0.11   \\
$\pi$ reconstruction            &1.4\\
$\eta$ reconstruction &4.1 \\
$K_S^0$ reconstruction &1.4 \\
PDF parametrization  &11.9 \\
PID selection &2.4 \\
Background suppression  &4.4 \\
\hline
Subtotal (without $f_s$)             &17.6   \\
\hline
$f_s$                           & 15.4  \\
\hline
Total                           & 23.4    \\
\hline \hline
\end{tabular}
\egroup
\label{tab:systematics}
\end{table}

To estimate the upper limit using the frequentist approach~\cite{Neyman}, 
an 80\% confidence-level (CL) belt (including systematic uncertainties) is prepared. 
To prepare this belt, we generate MC pseudoexperiments 
according to signal and background PDFs described previously. 
For each experiment, we generate 800 background events, 
which is, approximately, the number of background events obtained from fitting $\Upsilon(5S)$ data. 
We generate toy MC samples with the number of signal events in the range between 0 and 15. 
For each number of signal MC events we generate 2000 pseudoexperiments, 
obtain the number of signal events from a 3D fit 
and smear the the resulting distributions of signal yields 
using the Gaussian $\sigma$ of the total systematic uncertainty. 
The overall uncertainty 
is obtained by combining the uncertainty in the yield, $\sigma$, 
with the multiplicative uncertainty $\delta$ using the following 
formula~\cite{multiplicative}: 

\begin{equation}
(N \pm \sigma) (1\pm \delta) = N \pm (\sigma \oplus N\delta \oplus \sigma\delta), 
\end{equation}

\noindent where $\oplus$ denotes addition in quadrature. 
We use the results of our pseudoexperiments 
to prepare an 80\% classical confidence belt (without ordering), 
for which the lower and upper ends of respective confidence intervals  
correspond to the values for which 10\% of fitting results 
lie below and above the boundary of the contour. 
We use this 80\% confidence belt and its lower 10\% sideband to estimate 
a 90\% CL upper limit on the number of signal events in data 
to be 2.1, corresponding to a 90\% CL upper limit on the 
branching fraction $\mathcal{B}(B^0_s \rightarrow \eta^{\prime} K_S^0) <  8.16 \times 10^{-6}$.
We also estimate a 90\% CL upper limit 
on the product $f_s \times \mathcal{B}(\bepkk) < 1.64 \times 10^{-6}$.
The confidence intervals prepared using this statistical method 
are known to slightly ``overcover'' 
for the number of signal events~\cite{FeldmanCousins}, 
therefore resulting in a conservative upper limit. 


In summary, 
we search for the charmless rare decay $\bepkk$ 
using the full data sample collected by the Belle experiment at $\Upsilon(5S)$ resonance.
We find no statistically significant signal and set 90\% CL upper limits 
$\mathcal{B}(B^0_s \rightarrow \eta^{\prime} K_S^0) <  8.16 \times 10^{-6}$ 
and 
$f_s \times \mathcal{B}(\bepkk) < 1.64 \times 10^{-6}$.
Our results are the only experimental information currently available 
for this decay channel, 
and 
the reported 90\% CL upper limit on the branching fraction is 
several times larger than the current theoretical predictions 
based on QCDF, SCET and flavor SU(3) symmetry~\cite{bf1,bf2,bf3,bf4,bf5}. 
This decay should be further searched for 
by the Belle~II experiment~\cite{BelleII} 
at the next-generation $B$-factory SuperKEKB, 
where its discovery would require 
$\Upsilon(5S)$ statistics 
of the order of $2 {\rm \, ab^{-1}}$. 


We thank the KEKB group for the excellent operation of the
accelerator; the KEK cryogenics group for the efficient
operation of the solenoid; and the KEK computer group, and the Pacific Northwest National
Laboratory (PNNL) Environmental Molecular Sciences Laboratory (EMSL)
computing group for strong computing support; and the National
Institute of Informatics, and Science Information NETwork 5 (SINET5) for
valuable network support.  We acknowledge support from
the Ministry of Education, Culture, Sports, Science, and
Technology (MEXT) of Japan, the Japan Society for the 
Promotion of Science (JSPS), and the Tau-Lepton Physics 
Research Center of Nagoya University; 
the Australian Research Council including grants
DP180102629, 
DP170102389, 
DP170102204, 
DP150103061, 
FT130100303; 
Austrian Federal Ministry of Education, Science and Research (FWF) and
FWF Austrian Science Fund No.~P~31361-N36;
the National Natural Science Foundation of China under Contracts
No.~11435013,  
No.~11475187,  
No.~11521505,  
No.~11575017,  
No.~11675166,  
No.~11705209;  
Key Research Program of Frontier Sciences, Chinese Academy of Sciences (CAS), Grant No.~QYZDJ-SSW-SLH011; 
the  CAS Center for Excellence in Particle Physics (CCEPP); 
the Shanghai Science and Technology Committee (STCSM) under Grant No.~19ZR1403000; 
the Ministry of Education, Youth and Sports of the Czech
Republic under Contract No.~LTT17020;
Horizon 2020 ERC Advanced Grant No.~884719 and ERC Starting Grant No.~947006 ``InterLeptons'' (European Union);
the Carl Zeiss Foundation, the Deutsche Forschungsgemeinschaft, the
Excellence Cluster Universe, and the VolkswagenStiftung;
the Department of Atomic Energy (Project Identification No. RTI 4002) and the Department of Science and Technology of India; 
the Istituto Nazionale di Fisica Nucleare of Italy; 
National Research Foundation (NRF) of Korea Grant
Nos.~2016R1\-D1A1B\-01010135, 2016R1\-D1A1B\-02012900, 2018R1\-A2B\-3003643,
2018R1\-A6A1A\-06024970, 2019K1\-A3A7A\-09033840,
2019R1\-I1A3A\-01058933, 2021R1\-A6A1A\-03043957,
2021R1\-F1A\-1060423, 2021R1\-F1A\-1064008;
Radiation Science Research Institute, Foreign Large-size Research Facility Application Supporting project, the Global Science Experimental Data Hub Center of the Korea Institute of Science and Technology Information and KREONET/GLORIAD;
the Polish Ministry of Science and Higher Education and 
the National Science Center;
the Ministry of Science and Higher Education of the Russian Federation, Agreement 14.W03.31.0026, 
and the HSE University Basic Research Program, Moscow; 
University of Tabuk research grants
S-1440-0321, S-0256-1438, and S-0280-1439 (Saudi Arabia);
the Slovenian Research Agency Grant Nos. J1-9124 and P1-0135;
Ikerbasque, Basque Foundation for Science, Spain;
the Swiss National Science Foundation; 
the Ministry of Education and the Ministry of Science and Technology of Taiwan;
and the United States Department of Energy and the National Science Foundation.




%

\end{document}

%% file: pub605.tex
\noaffiliation
\affiliation{Department of Physics, University of the Basque Country UPV/EHU, 48080 Bilbao}
\affiliation{University of Bonn, 53115 Bonn}
\affiliation{Brookhaven National Laboratory, Upton, New York 11973}
\affiliation{Budker Institute of Nuclear Physics SB RAS, Novosibirsk 630090}
\affiliation{Faculty of Mathematics and Physics, Charles University, 121 16 Prague}
\affiliation{Chonnam National University, Gwangju 61186}
\affiliation{Chung-Ang University, Seoul 06974}
\affiliation{University of Cincinnati, Cincinnati, Ohio 45221}
\affiliation{Deutsches Elektronen--Synchrotron, 22607 Hamburg}
\affiliation{Duke University, Durham, North Carolina 27708}
\affiliation{Institute of Theoretical and Applied Research (ITAR), Duy Tan University, Hanoi 100000}
\affiliation{University of Florida, Gainesville, Florida 32611}
\affiliation{Key Laboratory of Nuclear Physics and Ion-beam Application (MOE) and Institute of Modern Physics, Fudan University, Shanghai 200443}
\affiliation{Justus-Liebig-Universit\"at Gie\ss{}en, 35392 Gie\ss{}en}
\affiliation{Gifu University, Gifu 501-1193}
\affiliation{II. Physikalisches Institut, Georg-August-Universit\"at G\"ottingen, 37073 G\"ottingen}
\affiliation{SOKENDAI (The Graduate University for Advanced Studies), Hayama 240-0193}
\affiliation{Gyeongsang National University, Jinju 52828}
\affiliation{Department of Physics and Institute of Natural Sciences, Hanyang University, Seoul 04763}
\affiliation{University of Hawaii, Honolulu, Hawaii 96822}
\affiliation{High Energy Accelerator Research Organization (KEK), Tsukuba 305-0801}
\affiliation{J-PARC Branch, KEK Theory Center, High Energy Accelerator Research Organization (KEK), Tsukuba 305-0801}
\affiliation{National Research University Higher School of Economics, Moscow 101000}
\affiliation{Forschungszentrum J\"{u}lich, 52425 J\"{u}lich}
\affiliation{IKERBASQUE, Basque Foundation for Science, 48013 Bilbao}
\affiliation{Indian Institute of Science Education and Research Mohali, SAS Nagar, 140306}
\affiliation{Indian Institute of Technology Guwahati, Assam 781039}
\affiliation{Indian Institute of Technology Hyderabad, Telangana 502285}
\affiliation{Indian Institute of Technology Madras, Chennai 600036}
\affiliation{Indiana University, Bloomington, Indiana 47408}
\affiliation{Institute of High Energy Physics, Chinese Academy of Sciences, Beijing 100049}
\affiliation{Institute of High Energy Physics, Vienna 1050}
\affiliation{Institute for High Energy Physics, Protvino 142281}
\affiliation{INFN - Sezione di Napoli, I-80126 Napoli}
\affiliation{INFN - Sezione di Roma Tre, I-00146 Roma}
\affiliation{INFN - Sezione di Torino, I-10125 Torino}
\affiliation{Iowa State University, Ames, Iowa 50011}
\affiliation{Advanced Science Research Center, Japan Atomic Energy Agency, Naka 319-1195}
\affiliation{J. Stefan Institute, 1000 Ljubljana}
\affiliation{Institut f\"ur Experimentelle Teilchenphysik, Karlsruher Institut f\"ur Technologie, 76131 Karlsruhe}
\affiliation{Kavli Institute for the Physics and Mathematics of the Universe (WPI), University of Tokyo, Kashiwa 277-8583}
\affiliation{Kitasato University, Sagamihara 252-0373}
\affiliation{Korea Institute of Science and Technology Information, Daejeon 34141}
\affiliation{Korea University, Seoul 02841}
\affiliation{Kyoto Sangyo University, Kyoto 603-8555}
\affiliation{Kyungpook National University, Daegu 41566}
\affiliation{Universit\'{e} Paris-Saclay, CNRS/IN2P3, IJCLab, 91405 Orsay}
\affiliation{P.N. Lebedev Physical Institute of the Russian Academy of Sciences, Moscow 119991}
\affiliation{Liaoning Normal University, Dalian 116029}
\affiliation{Faculty of Mathematics and Physics, University of Ljubljana, 1000 Ljubljana}
\affiliation{Ludwig Maximilians University, 80539 Munich}
\affiliation{Luther College, Decorah, Iowa 52101}
\affiliation{Malaviya National Institute of Technology Jaipur, Jaipur 302017}
\affiliation{Faculty of Chemistry and Chemical Engineering, University of Maribor, 2000 Maribor}
\affiliation{Max-Planck-Institut f\"ur Physik, 80805 M\"unchen}
\affiliation{School of Physics, University of Melbourne, Victoria 3010}
\affiliation{University of Mississippi, University, Mississippi 38677}
\affiliation{University of Miyazaki, Miyazaki 889-2192}
\affiliation{Moscow Physical Engineering Institute, Moscow 115409}
\affiliation{Graduate School of Science, Nagoya University, Nagoya 464-8602}
\affiliation{Kobayashi-Maskawa Institute, Nagoya University, Nagoya 464-8602}
\affiliation{Universit\`{a} di Napoli Federico II, I-80126 Napoli}
\affiliation{Nara Women's University, Nara 630-8506}
\affiliation{National Central University, Chung-li 32054}
\affiliation{Department of Physics, National Taiwan University, Taipei 10617}
\affiliation{H. Niewodniczanski Institute of Nuclear Physics, Krakow 31-342}
\affiliation{Nippon Dental University, Niigata 951-8580}
\affiliation{Niigata University, Niigata 950-2181}
\affiliation{University of Nova Gorica, 5000 Nova Gorica}
\affiliation{Novosibirsk State University, Novosibirsk 630090}
\affiliation{Osaka City University, Osaka 558-8585}
\affiliation{Pacific Northwest National Laboratory, Richland, Washington 99352}
\affiliation{Panjab University, Chandigarh 160014}
\affiliation{University of Pittsburgh, Pittsburgh, Pennsylvania 15260}
\affiliation{Punjab Agricultural University, Ludhiana 141004}
\affiliation{Research Center for Nuclear Physics, Osaka University, Osaka 567-0047}
\affiliation{Meson Science Laboratory, Cluster for Pioneering Research, RIKEN, Saitama 351-0198}
\affiliation{Department of Modern Physics and State Key Laboratory of Particle Detection and Electronics, University of Science and Technology of China, Hefei 230026}
\affiliation{Showa Pharmaceutical University, Tokyo 194-8543}
\affiliation{Soochow University, Suzhou 215006}
\affiliation{Soongsil University, Seoul 06978}
\affiliation{Sungkyunkwan University, Suwon 16419}
\affiliation{School of Physics, University of Sydney, New South Wales 2006}
\affiliation{Department of Physics, Faculty of Science, University of Tabuk, Tabuk 71451}
\affiliation{Tata Institute of Fundamental Research, Mumbai 400005}
\affiliation{Department of Physics, Technische Universit\"at M\"unchen, 85748 Garching}
\affiliation{Toho University, Funabashi 274-8510}
\affiliation{Department of Physics, Tohoku University, Sendai 980-8578}
\affiliation{Earthquake Research Institute, University of Tokyo, Tokyo 113-0032}
\affiliation{Department of Physics, University of Tokyo, Tokyo 113-0033}
\affiliation{Tokyo Institute of Technology, Tokyo 152-8550}
\affiliation{Tokyo Metropolitan University, Tokyo 192-0397}
\affiliation{Virginia Polytechnic Institute and State University, Blacksburg, Virginia 24061}
\affiliation{Wayne State University, Detroit, Michigan 48202}
\affiliation{Yamagata University, Yamagata 990-8560}
\affiliation{Yonsei University, Seoul 03722}
  \author{T.~Pang}\affiliation{University of Pittsburgh, Pittsburgh, Pennsylvania 15260} 
  \author{V.~Savinov}\affiliation{University of Pittsburgh, Pittsburgh, Pennsylvania 15260} 
  \author{I.~Adachi}\affiliation{High Energy Accelerator Research Organization (KEK), Tsukuba 305-0801}\affiliation{SOKENDAI (The Graduate University for Advanced Studies), Hayama 240-0193} 
  \author{H.~Aihara}\affiliation{Department of Physics, University of Tokyo, Tokyo 113-0033} 
  \author{D.~M.~Asner}\affiliation{Brookhaven National Laboratory, Upton, New York 11973} 
  \author{H.~Atmacan}\affiliation{University of Cincinnati, Cincinnati, Ohio 45221} 
  \author{V.~Aulchenko}\affiliation{Budker Institute of Nuclear Physics SB RAS, Novosibirsk 630090}\affiliation{Novosibirsk State University, Novosibirsk 630090} 
  \author{T.~Aushev}\affiliation{National Research University Higher School of Economics, Moscow 101000} 
  \author{R.~Ayad}\affiliation{Department of Physics, Faculty of Science, University of Tabuk, Tabuk 71451} 
  \author{V.~Babu}\affiliation{Deutsches Elektronen--Synchrotron, 22607 Hamburg} 
% ...
\collaboration{The Belle Collaboration}